\input epsf.tex
\documentclass[12pt]{article}
\usepackage{graphicx}
\usepackage{amssymb,epsf,amsmath}
\csname @addtoreset\endcsname{equation}{section}

\def\bseq{\begin{subequation}}  
\def\eseq{\end{subequation}}
\def\Bar#1{\overline{#1}}                       

\newcommand{\beq}{\begin{equation}}
\newcommand{\eeq}{\end{equation}}
\newcommand{\bea}{\begin{eqnarray}}
\newcommand{\eea}{\end{eqnarray}}
\newcommand{\ena}{\end{eqnarray}}

\newcommand {\non}{\nonumber}

\renewcommand{\a}{\alpha}
\renewcommand{\b}{\beta}

\renewcommand{\d}{\delta}
\renewcommand{\th}{\theta}

\newcommand{\pa}{\partial}
\newcommand{\g}{\gamma}
\newcommand{\G}{\Gamma}

\newcommand{\e}{\epsilon}

\renewcommand{\l}{\lambda}

\newcommand{\m}{\mu}

\newcommand{\f}{\phi}

\newcommand{\p}{\pi}

\newcommand{\Db}{\Bar{D}}

\newcommand{\Tr}{{\rm Tr}}

\newcommand{\intsupk}[1]{\int\!\! \frac{d^3{#1}}{(2\pi)^3} d^4\theta ~}

\begin{document}

\begin{titlepage}
{\hbox to\hsize{\hfill September 2010}}

\begin{center}
\vglue 0.99in
{\Large\bf  THE CONFORMAL MANIFOLD OF \\ [.09in]
CHERN--SIMONS MATTER THEORIES}
\\[.45in]
Marco S. Bianchi \footnote{marco.bianchi@mib.infn.it}  ~and~
Silvia Penati \footnote{silvia.penati@mib.infn.it} \\
{\it Dipartimento di Fisica dell'Universit\`a degli studi di
Milano-Bicocca,\\
and INFN, Sezione di Milano-Bicocca, piazza della Scienza 3, I-20126 Milano,
Italy}\\[.8in]

{\bf ABSTRACT}\\[.0015in]
\end{center}
 
We determine perturbatively the conformal manifold of ${\cal N}=2$  Chern--Simons matter theories with the aim of checking in the three dimensional case the general prescription based on global symmetry breaking,
recently introduced in \cite{Green:2010da,Kol:2010ub}. We discuss in details few remarkable cases like the ${\cal N}=6$ ABJM theory and its less supersymmetric generalizations with/without flavors.  In all cases we find perfect agreement with the predictions of  global symmetry breaking prescription.

${~~~}$ \newline

\end{titlepage}

\section{Introduction}

In this paper we determine the conformal manifold of ${\cal N}=2$ supersymmetric Chern--Simons matter theories in three dimensions, that is the space of perturbations which preserve conformality. This space is spanned by the exactly marginal deformations of the theory.

In general, two approaches can be used to determine the conformal manifold of a given theory. 
The first is based on the direct inspection of the $\b$--functions, requiring their vanishing. If this constraint can be satisfied imposing fewer conditions than the number of couplings one has a non--trivial manifold of conformal field theories. This method applies to all field theories which allow for a direct evaluation of the $\b$--functions.  In particular, this argument has been used to show that conformal manifolds are very common in four dimensional ${\cal N} = 1$ supersymmetric field theories, and has been applied to a wide class of examples \cite{Leigh:1995ep}.

The second approach is based on the study of global symmetry breaking and has been recently derived for ${\cal N} = 1$ theories in four dimensions \cite{Green:2010da,Kol:2010ub}. The main result is that the conformal manifold can be determined as the quotient ${\cal M} = \{\l\}/G_{\mathbb{C}}$ of the set of supermarginal operators of the theory by its complexified (continuous, non R--symmetry) global symmetry group.
This method is very powerful since it does not require the evaluation of $\b$--functions  and allows for the determination of the conformal manifold even for supersymmetric theories without a Lagrangian description. 
For several four dimensional examples it has been checked that the two methods agree \cite{Green:2010da,Kol:2010ub}. 

As discussed in \cite{Green:2010da}, nothing prevents the new approach from being applied to ${\cal N} = 2$ supersymmetric theories in three dimensions which have a  superspace description very similar to ${\cal N}=1$ theories in four dimensions.

During the past few years supersymmetric three dimensional theories have been intensively studied as candidates for describing the low energy effective theory living on M2 branes and for finding an explicit realization of $\text{AdS}_4/\text{CFT}_3$ correspondence \cite{schwarz}--\cite{Hosomichi:2008jd}.\\
After the seminal work by Aharony, Bergman, Jafferis, Maldacena \cite{ABJM}  who have determined the AdS dual theory of a three dimensional ${\cal N}=6$ Chern--Simons--matter CFT, a large class of supersymmetric three dimensional CFT's has been recently introduced \cite{Klebanov}--\cite{Chang:2010sg}.   
The ABJM theory is described by two gauge fields corresponding to two unitary groups and two pairs of bifundamental matter fields.   
A lagrangian description is available in terms of two ${\cal N}=2$ superspace Chern--Simons actions with opposite CS levels and a chiral action characterized by  a $SU(2)\times SU(2)$ invariant quartic superpotential \cite{Zupnik:1988en,Ivanov,Nishino:1991sr,Klebanov}.
Less supersymmetric  generalizations of this model include CFT's corresponding to different CS levels \cite{GT} or theories with ${\cal N}=2$ flavor degrees of freedom \cite{HK,GJ,HLT, BCC}.

It is the aim of this paper to apply the approach of \cite{Green:2010da,Kol:2010ub}  to the ABJM theory and its ${\cal N}=2$ generalizations for determining their conformal manifold and check it against a direct inspection of the $\b$--functions vanishing conditions.

We first apply the method of global symmetry breaking  \cite{Green:2010da,Kol:2010ub} to Chern--Simons matter 
theories characterized by different CS levels and  
determine their conformal manifold and the structure of the exactly marginal deformations (Section 2). We then check the result against a two--loop perturbative evaluation of both the dimensionality of the conformal manifold and the form of the exactly marginal deformations (Section 3). This goes through the determination of the two--loop beta functions for the most general ${\cal N}=2$ Chern--Simons matter theory obtained by adding the complete set of marginal superpotential terms \footnote{This analysis completes previous works on the deformations of the ABJM theory \cite{Bianchi:2009ja}, \cite{Bianchi:2009rf} where only a subset of marginal deformations was investigated.}.
Adding all possible perturbations induces a nontrivial operator mixing at two loops. Therefore,  the determination of an independent set of constraints for vanishing beta--functions requires diagonalizing the mixing matrix. Doing that, we eventually find a set of seven constraints for ten unknowns which leads to 
a three dimensional conformal manifold. Our perturbative results confirm the predictions of the general method based on global symmetry breaking. 

Finally, in Section 4 we generalize the analysis to ${\cal N}=2$  flavored theories.

\section{Exactly marginal deformations of Chern--Simons matter theories from global symmetry breaking}\label{sec:2}

We consider the class of Chern--Simons matter theories described by the ${\cal N} = 2$ superspace action  \cite{Zupnik:1988en,Ivanov,Nishino:1991sr,Klebanov}
\bea
\label{action}
{\cal S} &=& \int d^3x\,d^4\theta \int_0^1 dt\: \left\{ K_1 \Tr \Big[
  V \Db^\a \left( e^{-t V} D_\a e^{t V} \right) \Big]+ K_2   \Tr \Big[
  \hat{V} \Db^\a \left( e^{-t \hat{V}} D_\a
  e^{t \hat{V}} \right) \Big] \right\}+ \non \\
  && + \int d^3x\,d^4\theta\: \Tr \left( \bar{A}_a
  e^V A^a e^{- \hat{V}} + \bar{B}^a e^{\hat V} B_a
  e^{-V} \right)  ~+~ W 
\eea
where
\beq
W =    \frac{\l}{2}\, \int d^3x\,d^2\theta\: \,  \e_{ab}\,\e^{cd}\,
   \Tr  \left[ A^a B_c A^b B_d \right]  \, + \, h.c.
   \label{superpotential}
\eeq
Here $(V, \hat{V})$ are the gauge supermultiplets for $U(N) \times U(M)$ gauge group, whereas
the chiral multiplets $A^a$ and $B_a$, $a = 1,2$, belong to $(N,\bar{M})$
and $(\bar{N},M)$ representations, respectively. 

The two levels $(K_1, K_2)$ appearing in front of the Chern--Simons terms are integer parameters labeling different theories which are not connected by RG flow. The factors $1/K_1, 1/K_2$ measure the strength of the gauge interactions and in the limit of large $K_{1,2}$ a perturbative investigation is available.  
 
For $\l=0$ the classical global symmetry of the theory is $U(2)_A \times U(2)_B $ where the two groups act independently on the $A$ and $B$ doublets. The diagonal $U(1)$ is the baryonic symmetry under which the bifundamentals $\left( A^1,\, A^2,\, B_1,\, B_2 \right)$ have charges $(1,\,1,\,-1,\,-1)$. It is gauged into the $U(N)\times U(M)$ gauge group, and therefore it is conserved by any gauge invariant superpotential. The axial $U(1)$ acts on the bifundamental matter with charges $(1,\,1,\,1,\,1)$.
 
The addition of the marginal perturbation $\e_{ab}\,\e^{cd}\, \Tr  \left( A^a B_c A^b B_d \right)$ breaks the axial $U(1)$ and leads the theory to the infrared fixed point given by \cite{Bianchi:2009rf}
\beq
\l = \sqrt{ \frac{2MN +1}{2(MN-1)} \left( \frac{1}{K_1^2} + \frac{1}{K_2^2} \right) + \frac{MN+2}{MN-1} \frac{1}{K_1K_2} }
\label{fixedpt}
\eeq 
where only the $SU(2)_A \times SU(2)_B$ global symmetry survives. For $K_1 = -K_2 \equiv K$ it reduces to the ABJ(M) fixed point 
$\l = 1/K$ \cite{ABJM,ABJ} where supersymmetry gets enhanced to ${\cal N}=6$. 
  
In order to determine the structure of the exactly marginal deformations we have to find the class of gauge invariant supermarginal operators. For ${\cal N}=2$ supersymmetric theories in three dimensions classical marginal  perturbations are quartic in the bifundamental fields and constrained by gauge symmetry to have the form $\l_{ab}^{cd}\, \Tr(A^a B_c A^b B_d)$, where 
$\l_{ab}^{cd}$ is a tensor of $SU(2)\times SU(2)$.  For  $\l_{ab}^{cd} = 1/2\, \e_{ab}\, \e^{cd}$ 
we find the ABJ(M) superpotential which preserves the $SU(2)\times SU(2)$ global symmetry. In addition, there are $9$ operators transforming in the $(\mathbf{3},\mathbf{3})$ representation of $SU(2)\times SU(2)$ which have symmetric upper and lower indices, separately. Explicitly, they are
\bea
\label{marginal}
&& \Tr \left(A^1 B_1 A^2 B_2 + A^1 B_2 A^2 B_1 \right) , \\
&& \Tr \left(A^1 B_1 A^1 B_1 \right) ,\quad \Tr \left(A^2 B_2 A^2 B_2 \right) , \non\\
&& \Tr \left(A^1 B_2 A^1 B_2 \right) , \quad \Tr \left(A^2 B_1 A^2 B_1 \right) , \non\\
&& \Tr \left(A^1 B_1 A^1 B_2 \right) , \quad \Tr \left(A^1 B_1 A^2 B_1 \right) , \quad \Tr \left(A^1 B_2 A^2 B_2 \right) ,\quad \Tr \left(A^2 B_1 A^2 B_2 \right) 
\non 
\eea
All together they break $SU(2)\times SU(2) \times U(1)_{axial}$ completely.

The gauge sector has no marginal operators since the Chern--Simons term is not gauge invariant. On the other hand, Yang--Mills contributions cannot be added, being them dimensionful. 

In order to investigate the structure of the conformal manifold for this kind of theories, we perturb the action (\ref{action}) at its conformal point (\ref{fixedpt}) by adding the marginal chiral operators (\ref{marginal}). In the class of marginal perturbations we do not include the ABJM superpotential $\l_{ab}^{cd}\, \Tr(A^a B_c A^b B_d)$ since at the IR fixed point 
it is irrelevant. In fact, since this operator leads the theory from the UV fixed point (free theory) to the nontrivial IR point, it is 
marginal but not exactly marginal. As discussed in \cite{Green:2010da}, it becomes irrelevant by pairing with the 
short current multiplet associated to the axial $U(1)$ global symmetry broken by the operator itself. This will be checked perturbatively in Section \ref{sec:OPE}.

According to the general prescription of  \cite{Green:2010da,Kol:2010ub}, the conformal manifold is given locally by the symplectic quotient $\{\l\}/G_{\mathbb{C}}$, where $\{\l\}$ is the set of all marginal perturbations of the CFT and $G_{\mathbb{C}}$ is the complexified
global (continuous) symmetry group.  

In our case $\{ \l \}$ is given by the set (\ref{marginal}), whereas the global symmetry group at the IR fixed point is $SU(2) \times SU(2)$. We can thus write
\bea
{\cal M}_c = \frac{ (\mathbf{3},\mathbf{3})}{(SU(2)\times SU(2))_{\mathbb{C}}}
\label{quotient}
\eea
and the complex dimension of the conformal manifold is $9-6=3$. This means that there are $3$ exactly marginal operators which we now identify.

First of all, the most general linear combination of marginal chirals (\ref{marginal}) can be written as the matrix product \cite{Benvenuti:2005wi}
\bea
W = M_{ij} w_i w_j
\eea
where the vectors $w$ are defined as
\bea
w_1 &=& \frac12(A^1 B_1 + A^2 B_2)\non\\
w_2 &=& \frac{i}{2}(A^1 B_1 - A^2 B_2) \non\\
w_3 &=& \frac{i}{2}(A^1 B_2 + A^2 B_1)\non\\
w_4 &=& \frac12 (A^1 B_2 - A^2 B_1)
\eea
and $M$ is a symmetric traceless $4\times 4$ matrix. The combinations which survive the quotient (\ref{quotient}) are obviously the diagonal ones. Out of them we can find three independent chiral primary operators 
\bea\label{exactlymarginal}
&& \Tr \left( A^1 B_1 A^1 B_1 + A^2 B_2 A^2 B_2 \right) \non\\
&& \Tr \left( A^1 B_1 A^2 B_2 + A^1 B_2 A^2 B_1 \right)  \non\\
&& \Tr \left( A^1 B_2 A^1 B_2 + A^2 B_1 A^2 B_1 \right) 
\eea
Therefore,  including the original ABJM superpotential term, we conclude that the exactly marginal superpotential should have the form
\bea\label{exactlymarginalsp}
&& u\, \Tr \left( A^1 B_1 A^1 B_1 + A^2 B_2 A^2 B_2 \right) + f(t,u,v)\, \Tr \left( A^1 B_1 A^2 B_2 - A^1 B_2 A^2 B_1 \right)  \non\\
&& t\, \Tr \left( A^1 B_1 A^2 B_2 + A^1 B_2 A^2 B_1 \right) + v\, \Tr \left( A^1 B_2 A^1 B_2 + A^2 B_1 A^2 B_1 \right) 
\eea
where $f$ is a solution of the D--term conditions \cite{Green:2010da}.
The explicit form of the function $f(t,u,v)$ which describes the conformal manifold will be determined at two--loops in the next Section.

\vskip 10pt
The same analysis can be applied to other remarkable fixed points introduced in \cite{GT} and perturbatively determined in 
\cite{Bianchi:2009ja,Bianchi:2009rf}. 

We consider a ${\cal N}=2$ theory described by the action (\ref{action}) with superpotential
\beq
W =      \int d^3x\,d^2\theta\: \,  
   \Tr  \left[  c_1 (A^a B_a)^2 + c_2  (B_a A^a)^2 \right]  \, + \, h.c.
   \label{superpotential2}
\eeq
For real couplings, the equation
\beq
c_1^2 + c_2^2 + 2 \frac{MN+2}{2MN+1} c_1 c_2 = \frac{1}{4 K_1^2} + \frac{1}{4 K_2^2} 
+ \frac{MN+2}{2 K_1 K_2 (2MN +1)}
\eeq
describes a line of fixed points \cite{Bianchi:2009rf}. For 
particular values $c_1 = \frac{1}{2K_1}$ and $c_2 = \frac{1}{2K_2}$ supersymmetry gets enhanced to ${\cal N}=3$ \cite{GT}. 

We focus on a given point on this line and study the conformal manifold perturbing around it with all classical marginal operators. 
This time the superpotential preserves a diagonal global $SU(2)$ subgroup, out of the original 
$SU(2)\times SU(2)\times U(1)_{axial}$
global symmetry of the free theory. Therefore, four classically marginal operators have become irrelevant at the IR fixed point by coupling to the broken currents of the four broken generators. The set of supermarginals is thus reduced to six operators. The dimension of the conformal manifold is then given by
\bea
\text{dim}\left( \frac{\{6\,\, \text{operators}\}}{SU(2)_{\mathbb{C}}}\right) = 6 - 3 =3
\eea

\section{Exactly marginal deformations of Chern--Simons matter theories from perturbation theory}\label{sec:3}

In this section we check the above statements on the structure of the conformal manifold against a perturbative calculation. 
As a byproduct, we find the explicit expression of the exactly marginal superpotential (\ref{exactlymarginalsp}).
 
We consider the most general action (\ref{action}) with superpotential  
\bea\label{Wgeneral}
W &=&  \, {\rm Tr} \Big[~ h_1 \, (A^1 B_1 A^2 B_2 - A^1 B_2 A^2 B_1)  + h_2 \, (A^1 B_1 A^2 B_2 + A^1 B_2 A^2 B_1)  
\\&& ~~~+ h_3\, A^1 B_1 A^1 B_1  + h_4\, A^2 B_2 A^2 B_2 + h_5\, A^1 B_2 A^1 B_2 + h_6\, A^2 B_1 A^2 B_1  
\non\\
&& ~~~+\, h_7\, A^1 B_1 A^1 B_2 + h_8\, A^1 B_1 A^2 B_1 + h_9\, A^2 B_2 A^2 B_1 + h_{10}\, A^2 B_2 A^1 B_2 \Big]  \, + \, h.c.
\non
\eea
which includes all classically supermarginal deformations.  This can be thought as the most general marginal perturbation of any of the CFT theories investigated in the previous Section.  

In order to determine perturbatively the conformal manifold of the theory we need evaluate its renormalization group functions. We do it at the first non--trivial order, that is two loops \cite{KaoLee,Kazakov,KLL}
, using the ${\cal N} = 2$ superspace quantization \cite{Nishino:1991sr} described in \cite{Bianchi:2009rf}. This leads to the following superspace propagators for the gauge sector (in Landau gauge)
\bea
  \langle V^A(1) \, V^B(2) \rangle
   = -\frac{1}{K_1} \frac{1}{\Box}  \Db^\a D_\a  \, \delta^4(\th_1-\th_2) \, \delta^{AB} \\ \langle \hat
   V^A(1) \, \hat V^B(2) \rangle =
   -\frac{1}{K_2} \frac{1}{\Box}  \Db^\a D_\a \,  \delta^4(\th_1-\th_2) \, \delta^{AB}
\label{gaugeprop}
\eea
and  
\bea
&&\langle \left(\bar A_a \right)^{\hat m}_{\ m}(1) \, \left(A^b \right)^n_{\ \hat n}(2) \rangle
  = -\frac{1}{\Box} \delta^4(\th_1 - \th_2) \, \delta^{\hat m}_{\ \hat
  n} \, \delta^{\ n}_{m} \, \delta^{\ b}_{a}
  \\
&&  \langle \left(\bar B^a \right)^m_{\ \hat m}(1) \, \left(B_b \right)^{\hat n}_{\ n}(2) \rangle = 
  -\frac{1}{\Box} \delta^4(\th_1 - \th_2) \, \delta^m_{\ n} \, \delta^{\ \hat n}_{\hat m} \delta^{a}_{\ b}
\eea
for the matter sector, while interaction vertices can be easily inferred from the expansion of the action (\ref{action}, \ref{Wgeneral}). 
 
\begin{figure}
  \center
\includegraphics[width = 0.75 \textwidth]{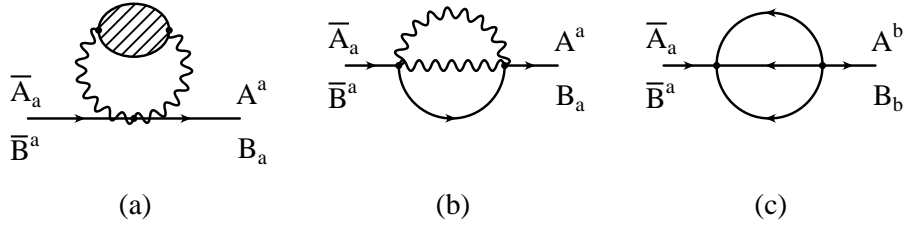}
  \caption{ Two--loop divergent diagrams contributing to the matter propagators.}
  \label{fig:diagrams}
\end{figure}

The only divergent graphs are self-energy diagrams for the bifundamental fields, which then contribute to their anomalous dimensions. They are given in Fig. \ref{fig:diagrams} where the bubble in the first diagram indicates the finite one--loop correction to the vector propagators \cite{Bianchi:2009rf} which in momentum space reads
\bea
&&  \Pi^{(1)} = \left[ - \frac{1}{8} f^{ABC} f^{A^\prime BC} + M \delta^{AA^\prime} \right] \intsupk{p} B_0(p)\,
  V^A(p)\, \Db^\a D^2 \Db_\a \, V^{A^\prime}(-p)
\non \\
&&  \hat \Pi^{(1)} = \left[ - \frac{1}{8} {\hat f}^{ABC} {\hat f}^{A^\prime BC} + N \delta^{AA^\prime} \right] \intsupk{p} B_0(p)\,
   {\hat V}^A(p)\, \Db^\a D^2 \Db_\a \, {\hat V}^{A^\prime}(-p) 
  \non \\
&&  \tilde \Pi^{(1)} = -2 \sqrt{NM} \delta^{A 0} \delta^{A^\prime
  0} \intsupk{p} B_0(p)\, V^A(p) \, \Db^\a D^2 \Db_\a \, \hat
  V^{A^\prime}(-p)  
\eea
Here $B_0(p) = 1/(8|p|)$ is the three dimensional bubble scalar integral.

The divergent contributions to the effective action are evaluated by dimensional regularization ($D= 3 -2\e$) and lead to
\bea\label{eq:efflag}
 \G_{2loops} &\rightarrow&  ~~  \left( 1 + \frac{1}{\e} d^A_{11} \right) \overline{A}_1 A^1+    \left( 1 +  \frac{1}{\e} d^A_{22} \right)
 \overline{A}_2 A^2 
\\&&
 +  \left( 1 +  \frac{1}{\e} d^B_{11} \right) \overline{B}^1 B_1+   \left( 1 + \frac{1}{\e}  d^B_{22} \right) \overline{B}^2 B_2 
 \non\\&&
+  ~  \frac{1}{\e}  d^A_{12} \; \overline{A}_1 A^2  +   \frac{1}{\e}  d^A_{21}  \; \overline{A}_2 A^1 +
 \frac{1}{\e}  d^B_{12} \;  \overline{B}^1 B_2   + \frac{1}{\e}  d^B_{21} \;  \overline{B}^2 B_1
 \non
\eea
where, calling 
\bea
C \equiv 2 \, MN \left( |h_1|^2 + |h_2|^2 \right) + 2\, \left( |h_2|^2 - |h_1|^2 \right) - \left( 2MN + 1 \right) \left(\frac{1}{K_1^2} + \frac{1}{K_2^2}  \right) - 2 \frac{MN + 2}{K_1 K_2} \non
\eea
the coefficients of the $\e$--poles in the $A$ sector are given by
\bea
d^A_{11} &=&  \frac{1}{64\,
\pi^2} \left\{C+ \left( MN +1 \right)\left[ 4 \left( |h_3|^2 + |h_5|^2 \right) + 2 |h_7|^2 + |h_8|^2 + |h_{10}|^2 \right]\right\}\non\\
d^A_{21} &=& \overline{d^A_{12}}  ~=~ \frac{1}{32\,
\pi^2} \left\{ h_2\, \overline{h_9} + h_7\, \overline{h_2} + h_3 \,\overline{h_8} +  h_5\, \overline{h_{10}} +  h_8 \,\overline{h_6} +  h_{10}\, \overline{h_4}  \right\} \non\\
d^A_{22} &=& \frac{1}{64\,
\pi^2} \left\{C+ \left( MN +1 \right)\left[ 4 \left( |h_4|^2 + |h_6|^2 \right) + 2 |h_9|^2 + |h_8|^2 + |h_{10}|^2 \right]\right\}
\eea
while the ones in the $B$ sector are obtained by replacing 
$h_5 \leftrightarrow h_6, \, h_7 \leftrightarrow h_8,\;h_9 \leftrightarrow h_{10}$. 
We note that the couplings $h_7$ -- $h_{10}$ cause mixing both in the $A$ and $B$ sectors due to non--vanishing $d^A_{21}$
and $d^B_{21}$ contributions.  

We renormalize the theory by adding suitable counterterms and defining 
\bea\label{eq:renfields}
A^a = \left( Z_{ab}^A \right)^{-\frac12} A^b_{\,0}    \qquad , \qquad B_a = \left( Z_{ab}^B \right)^{-\frac12} B_{b\,0}
\eea
in such a way that the bare quadratic action goes back being diagonal
\bea\label{eq:CT}
 {\cal L}^{(2)} + {\cal L}^{ct} &= &   ~~ \left( 1 - \frac{1}{\e} d^A_{11} \right) \overline{A}_1 A^1+    \left( 1 -  \frac{1}{\e} d^A_{22} \right)
 \overline{A}_2 A^2 -   \frac{1}{\e}  d^A_{12} \; \overline{A}_1 A^2  -   \frac{1}{\e}  d^A_{21}  \; \overline{A}_2 A^1 
\non \\
&&
+   \left( 1 -  \frac{1}{\e} d^B_{11} \right) \overline{B}^1 B_1+   \left( 1 - \frac{1}{\e}  d^B_{22} \right) \overline{B}^2 B_2 -  \frac{1}{\e}  d^B_{12} \;  \overline{B}^1 B_2   - \frac{1}{\e}  d^B_{21} \;  \overline{B}^2 B_1
 \non\\ 
 &&~~
 \non \\
 &=&  ~~\overline{A}_{1\,0}\, A^1_{\,0} + \overline{A}_{2\,0}\, A^2_{\,0} + \overline{B}^1_{\,0}\, B_{1\,0} + \overline{B}^2_{\,0}\, B_{2\,0} 
\eea
This requires the matrices $Z$ to be
\bea
Z^A = {\cal I}_{2\times2} - \frac{1}{\e}\left(\begin{array}{cc} d^A_{11} & d^A_{12} \\  \overline{d^A_{12}} & d^A_{22} \end{array}\right) \qquad
Z^B = {\cal I}_{2\times2} - \frac{1}{\e}\left(\begin{array}{cc} d^B_{11} & d^B_{12} \\  \overline{d^B_{12}} & d^B_{22} \end{array}\right)
\eea
The matrices of anomalous dimensions are defined as
\bea
  \g^A_{ab} \equiv \frac12 \frac{\pa \log{Z^A_{ab}}}{\pa \log{\mu}} \qquad\qquad  \g^B_{ab} \equiv \frac12 \frac{\pa \log{Z^B_{ab}}}{\pa \log{\mu}}
\label{eqn:anomdim}
\eea
and can be evaluated explicitly by
\bea\label{eq:defanomalous}
\g^A_{ab} = \sum_k \, h_k \, \frac{\pa d^A_{ab}}{\pa h_k}  \qquad\qquad  \g^B_{ab} = \sum_k \, h_k \, \frac{\pa d^B_{ab}}{\pa h_k}
\eea
Since the divergences are all quadratic in the couplings, this yields to
\bea\label{eq:anomalous}
\g^A_{11} & = & \frac{1}{32\,\pi^2}\left\{C+ \left( MN +1 \right)\left[ 4 \left( |h_3|^2 + |h_5|^2 \right) + 2 |h_7|^2 + |h_8|^2 + |h_{10}|^2 \right]\right\} \non\\
\g^A_{21} & = &  \overline{\g^A_{12}} ~=~  \frac{1}{16\, \pi^2}\left\{ h_2\, \overline{h_9} + h_7\, \overline{h_2} + h_3 \,\overline{h_8} +  h_5\, \overline{h_{10}} +  h_8 \,\overline{h_6} +  h_{10}\, \overline{h_4}  \right\} 
  \non\\
\g^A_{22} & = & \frac{1}{32\,\pi^2}\left\{C+ \left( MN +1 \right)\left[ 4 \left( |h_4|^2 + |h_6|^2 \right) + 2 |h_9|^2 + |h_8|^2 + |h_{10}|^2 \right]\right\}
\non \\
\g^B_{11} & = & \frac{1}{32\,\pi^2}\left\{C+ \left( MN +1 \right)\left[ 4 \left( |h_3|^2 + |h_6|^2 \right) + 2 |h_8|^2 + |h_7|^2 + |h_{9}|^2 \right]\right\} \non\\
\g^B_{21} & = &  \overline{\g^B_{12}} ~=~  \frac{1}{16\, \pi^2}\left\{ h_2\, \overline{h_{10}} + h_8\, \overline{h_2} + h_3 \,\overline{h_7} +  h_6\, \overline{h_{9}} +  h_7 \,\overline{h_5} +  h_{9}\, \overline{h_4}  \right\} 
  \non\\
\g^B_{22} & = & \frac{1}{32\,\pi^2}\left\{C+ \left( MN +1 \right)\left[ 4 \left( |h_4|^2 + |h_5|^2 \right) + 2 |h_{10}|^2 + |h_7|^2 + |h_{9}|^2 \right]\right\}
\eea
The  nonrenormalization theorem  for ${\cal N}=2$ supersymmetric CS theories in three dimensions \cite{Chang:2010sg} prevents 
the appearance of divergent corrections to the superpotential which then gets modified only by field function  
renormalization. Therefore, we renormalize the couplings as 
\beq
h_i = \m^{-2\e}\, Z_{h_i}^{-1}\, h_{i\,0} \qquad {\rm with } \qquad Z_{h_i} = \prod_{\Phi_i} Z^{-\frac12}_{\Phi_i}
\eeq
where $\Phi_i \equiv (A,\,B)$.
As a consequence,  the beta--functions
$\b_i = \mu \frac{d h_i}{d \mu}$ turn out to be expressed directly in terms of the coefficients of the anomalous dimensions
matrices. The calculation is straightforward and we find
\bea\label{eq:beta}
\b_{h_1} &=& h_1 \left( \g^A_{11} + \g^B_{11} + \g^A_{22} + \g^B_{22} \right) \non\\
\b_{h_2} &=& h_2 \left( \g^A_{11} + \g^B_{11} + \g^A_{22} + \g^B_{22} \right) + \left( h_7\, \overline{\g^A_{12}} + h_8\, \overline{\g^B_{12}} + h_9\, \g^A_{12} + h_{10}\, \g^B_{12} \right) \non\\
\b_{h_3} &=& 2\, h_3 \left( \g^A_{11} + \g^B_{11} \right) + h_7\, \g^B_{12} + h_8\, \g^A_{12}  \non\\
\b_{h_4} &=& 2\, h_4 \left( \g^A_{22} + \g^B_{22} \right) + h_9\, \overline{\g^B_{12}} + h_{10}\, \overline{\g^A_{12}} \non\\
\b_{h_5} &=& 2\, h_5 \left( \g^A_{11} + \g^B_{22} \right) + h_7\, \overline{\g^B_{12}} + h_{10}\, \g^A_{12}  \non\\
\b_{h_6} &=& 2\, h_6 \left( \g^A_{22} + \g^B_{11} \right) + h_8\, \overline{\g^A_{12}} + h_9\, \g^B_{12}  \non\\
\b_{h_7} &=& h_7 \left( 2 \g^A_{11} + \g^B_{11} + \g^B_{22} \right) + 2\left(  h_3\, \overline{\g^B_{12}} + h_2\, \g^A_{12} + h_5\, \g^B_{12} \right)\non\\
\b_{h_8} &=& h_8 \left( \g^A_{11} + \g^A_{22} + 2 \g^B_{11} \right) + 2\left(  h_3\, \overline{\g^A_{12}} + h_2\, \g^B_{12} + h_6\, \g^A_{12} \right)\non\\
\b_{h_9} &=& h_9 \left( 2 \g^A_{22} + \g^B_{11} + \g^B_{22} \right) + 2\left(  h_4\, \g^B_{12} + h_2\, \overline{\g^A_{12}} + h_6\, \overline{\g^B_{12}} \right)\non\\
\b_{h_{10}} &=& h_{10} \left( \g^A_{11} + \g^A_{22} + 2 \g^B_{22} \right) + 2\left(  h_4\, \g^A_{12} + h_2\, \overline{\g^B_{12}} + h_5\, \overline{\g^A_{12}}\right)
\eea
We see that due to the mixing of the anomalous dimensions, the $\b$--functions are not diagonal in the coupling constants.

The set of fixed points of the theory, that is its conformal manifold,  is determined by setting $\b_{h_i} =0$. In order to work out these constraints, it is convenient to diagonalize the $\b$--functions (\ref{eq:beta}).  In terms of the eigenvectors $h'_i$ the diagonal $\b$--functions read $\b'_{h_i} = \l_i h'_i$, where the eigenvalues $\l_i$ are 
\bea
\l_1 &=& \l_2 \equiv \l \non\\
\l_{3,4} &=& \l \pm \sqrt{\m_A}\non\\
\l_{5,6} &=& \l \pm \sqrt{\m_B}\non\\
\l_{7,8,9,10} &=& \l \pm \sqrt{\m_A + \m_B \pm 2 \sqrt{\m_A \m_B}}\non
\eea
where
\bea
 &\qquad& \qquad \qquad  \l = \g^A_{11} + \g^A_{22} + \g^B_{11} + \g^B_{22}\\
\m_A &=& (\g^A_{11} - \g^A_{22})^2 + 4 \g^A_{12} \overline{\g^A_{12}}\qquad , \qquad 
\m_B = (\g^B_{11} - \g^B_{22})^2 + 4 \g^B_{12} \overline{\g^B_{12}}
\non
\eea
Therefore, the set of conditions $\b_{h_i}=0$ is equivalent to the set $\l_i=0$  which implies
\beq\label{eq:gamma}
\g^A_{11} = \g^A_{22} \quad , \quad  \g^B_{11} = \g^B_{22}  \quad , \quad \g^A_{12} = \g^B_{12} = 0
\quad , \quad  \g^A_{11}   + \g^B_{11}   = 0 
\eeq
Since the diagonal coefficients of the anomalous dimensions matrices are real, while the off-diagonal ones are complex (see 
eq. (\ref{eq:anomalous})), we have obtained seven real conditions. In addition, it is easy to see that by a field redefinition we can remove the phases of seven couplings. All together there are seven complex contraints to be imposed on ten complex coupling constants. According to \cite{Leigh:1995ep}, this means that the conformal manifold of ${\cal N}=2$ Chern--Simons matter theories has complex dimension three.
In particular, for $K_1 = -K_2$ in (\ref{action}) this is the manifold of conformal fixed points connected to the ABJ(M) model by exactly marginal deformations.\\  This result matches the dimensions computed from the quotient $\{\l\}/G_{\mathbb{C}}$.

\vskip 10pt
We look for solutions of the constraints (\ref{eq:gamma}) in the space of the coupling constants in order
to determine the expression of the exactly marginal perturbations. 

The simplest case is obtained by setting $h_i=0$ $\forall i > 6$. This corresponds to the class of models investigated in \cite{Bianchi:2009rf}. At this locus the off--diagonal elements of the anomalous dimensions matrices vanish. The explicit solution to (\ref{eq:gamma}) in terms of the remaining coupling constants reads
\bea
\label{solution}
&& |h_3| = |h_4|,\quad |h_5| = |h_6|, \non\\&& (MN+1) \left[ |h_2|^2 + 2 |h_3|^2 + 2 |h_5|^2 \right] + (MN-1) |h_1|^2 = X 
\eea
where $X$ is coupling independent and given by
\bea
X = \frac{MN+2}{K_1K_2} + \frac{2MN+1}{2}\left(\frac{1}{K_1^2}+\frac{1}{K_2^2}\right) \non
\eea
Eq. (\ref{solution}) describes a 3--dimensional ellipsoid. In fact, by a field redefinition we can fix three phases: 
Setting $h_i = |h_i| e^{i\f_i}$,  we can for instance choose $\f_1=0$, $\f_{3}=\f_{4}$ and $\f_{5}=\f_{6}$.
It follows that the conformal manifold is parameterized by three complex parameters, for instance $h_2$, $h_3$ and $h_5$.  Its parametric equations may be written as
\bea
\label{solution2}
&& h_2 = t , \quad h_3 = h_4 = u ,\quad    h_5 = h_6 = v,   \non\\
&& |h_1|^2 =  \frac{1}{MN-1} \left\{ X - (MN+1) \left[ |t|^2 + 2 |u|^2 + 2 |v|^2 \right]\right\} \non\\
&& h_7 = h_8 = h_9 = h_{10} = 0
\eea
The reality of the $h_1$ solution requires the $(t,\,u,\,v)$ moduli to be not arbitrarily large, being subject
to the condition
\bea
 \left[ |t|^2 + 2 |u|^2 + 2 |v|^2 \right] \le 
\frac{MN+2}{MN+1} \frac{1}{K_1K_2} + \frac{2MN+1}{2(MN+1)}\left(\frac{1}{K_1^2}+\frac{1}{K_2^2}\right)
\eea
Since the limiting value is of order $1/K_{1,2}$, this condition also guarantees that in the large $K_{1,2}$ limit the results are within the range of the perturbative analysis. 
 
We have found that the conformal manifold is a compact surface for any ${\cal N}=2$ CS matter theory in three dimensions. In particular, it is true for the ABJ(M) theory and for $SU(2)$ invariant models, including the ${\cal N} = 3$ theory.\\ This renders the three dimensional models striking different from analogous models in four dimensions, where the exactly marginal couplings are not usually constrained \cite{Leigh:1995ep}.

The solution (\ref{solution2}) leads to an exactly marginal superpotential  of the form (\ref{exactlymarginalsp})
with $f(t,u,v)$ explicitly determined at two loops as
\bea
W_{ex\,mar}\!\! &=& \!\! \Tr \left[ t \left( A^1 B_1 A^2 B_2 + A^1 B_2 A^2 B_1 \right) + u \left( A^1 B_1 A^1 B_1 + A^2 B_2 A^2 B_2 \right) \right. \non\\\!\!&&\!\! + 
\frac{1}{\sqrt{MN-1}}\sqrt{X - (MN+1) \left( |t|^2 + 2 |v|^2 + 2 |u|^2 \right)} \left( A^1 B_1 A^2 B_2 - A^1 B_2 A^2 B_1 \right) 
\non\\
\!\!&&\!\! \left. +  v \left( A^1 B_2 A^1 B_2 + A^2 B_1 A^2 B_1 \right) \right] \, + \, h.c.
\eea
Locally, at the ABJ(M) fixed point the exactly marginal directions are given by a basis of operators on the tangent space with respect to the conformal surface ($ \partial_{t,u,v} h_i $), which corresponds to the operators
\bea
\label{eq:exmargop}
&& \Tr \left( A^1 B_1 A^1 B_1 + A^2 B_2 A^2 B_2 \right) \non\\
&& \Tr \left( A^1 B_1 A^2 B_2 + A^1 B_2 A^2 B_1 \right)  \non\\
&& \Tr \left( A^1 B_2 A^1 B_2 + A^2 B_1 A^2 B_1 \right) 
\eea
whereas the ABJM operator, which is transverse to the surface, is an irrelevant perturbation, matching the statement \cite{Green:2010da} that marginal deformations of such SCFT's can be either exactly marginal or marginally irrelevant, but never relevant. The same holds for all points belonging to the conformal manifold, which are therefore IR attractors.

As mentioned before, different solutions to the equations (\ref{eq:gamma}) may be found, corresponding to different conformal manifolds. For instance, we can find another set of solutions by setting  $h_3 = h_4 = h_5 = h_6 = 0$. It is easy to see that, again, we obtain a complex three dimensional manifold. It is related to the previous one by a $SO(4)$ rotation of the superpotential.  Since this is in general the case, the exactly marginal deformations (\ref{exactlymarginal}) are to be regarded as representatives of the equivalence classes of the quotient. 

\vskip 10pt
Finally, we prove that all the theories belonging to the conformal manifold are two--loop finite, having vanishing anomalous dimensions.
Indeed, from (\ref{eq:gamma}) we read that on the conformal manifold $\g^A_{11}=\g^A_{22}$ and $\g^B_{11}=\g^B_{22}$. Using 
the explicit expressions for the anomalous dimensions given in (\ref{eq:anomalous}), these constraints
give
\bea
4 \left( |h_3|^2 - |h_4|^2 \right) = |h_9|^2 + |h_{10}|^2 - |h_7|^2 - |h_8|^2 \non\\
4 \left( |h_5|^2 - |h_6|^2 \right) = |h_8|^2 + |h_9|^2 - |h_7|^2 - |h_{10}|^2 \non
\eea
Plugging these constraints into the two--loop anomalous dimensions (\ref{eq:anomalous}) we obtain that not only the matrices of anomalous dimensions are proportional to the identity, but also $\g^A_{11}=\g^A_{22}=\g^B_{11}=\g^B_{22} = \g$. 
On the other hand, the last condition in (\ref{eq:gamma}) states that their sum $\g^A_{11}+\g^A_{22}+\g^B_{11}+\g^B_{22} = 4\g$ should be zero, meaning that all anomalous dimensions are vanishing. We conclude that all theories belonging to the conformal manifold are two--loop finite.
A special case is the theory obtained by turning off the superpotential. This model is trivially scale invariant since there are no running couplings. However the anomalous dimensions are non--vanishing, because the bifundamental fields have to be renormalized due to divergent contributions to the effective action coming from gauge interactions. This theory is the UV fixed point of the RG flow. In particular the anomalous dimensions are strictly negative at this fixed point, as can be inferred from (\ref{eq:anomalous}). This means that adding a small superpotential deformation $h {\cal O}$ the operator ${\cal O}$ is relevant and will drive the theory to an IR fixed point, which is precisely what happens for the ABJ(M) model and the other ${\cal N} = 2$ CFT's in general.

\section{Currents and OPE}\label{sec:OPE}

In this Section we study the OPE algebra of marginal chiral operators and global currents at the first nontrivial perturbative order.
We show that the constraints for vanishing $\b$--functions previously found are in one to one correspondence with those for the vanishing of the D--terms for global symmetries \cite{Green:2010da,Kol:2010ub}.

Given the set of marginal chiral primary operators $\{ {\cal O}_i \}$ as appearing in the superpotential (\ref{Wgeneral}) (${\cal O}_i$ is the operator associated  to $h_i$) the OPE algebra has the general structure 
\bea\label{eq:OPE}
{\cal O}_i(x) \, \overline{{\cal O}_j}(0) = \frac{g_{i\overline{j}}}{|x|^4} + \frac{T^a_{i\overline{j}}}{|x|^3}\,J_a + \dots
\eea
where $g_{i\overline{j}}$ is the Zamolodchikov metric and $J_a$  the global symmetry currents.
Using the chiral propagators in coordinate space
\beq
\langle A^a(x) \bar{A}_b(0) \rangle =  \frac{1}{4\,\p}\, \frac{1}{|x|} \, \d^a_b \qquad , \qquad 
\langle B_a(x) \bar{B}^b(0) \rangle =  \frac{1}{4\,\p}\, \frac{1}{|x|} \, \d_a^b 
\eeq
it is easy to determine the metric on the couplings space at the lowest perturbative order 
\bea
g_{i\overline{j}} = \frac{ MN (MN+1)}{(4\pi)^4} \, \text{diag} \left( 2\,\frac{1-MN}{MN+1},\, 2 ,\, 2 ,\, 2,\, 2 ,\, 2 ,\, 1,\, 1,\, 1,\, 1 \right)
\eea
We note that it turns out to be diagonal, thanks to the particular choice of ${\cal O}_i$ in (\ref{Wgeneral}).

Performing three contractions we obtain the contributions of order $1/|x|^3$. They turn out to be expressed in terms of the global symmetries currents
\bea
&& J_{U(1)} = \overline{A}_1 A^1 + \overline{A}_2 A^2 + \overline{B}^1 B_1 + \overline{B}^2 B_2 \non\\
J^A_{+} &=&   \overline{A}_1 A^2 \quad , \quad 
J^A_{-} = \overline{A}_2 A^1 \quad , \quad ~
J^A_3 = \overline{A}_1 A^1 - \overline{A}_2 A^2 \non\\
J^B_{+} &=&  \overline{B}^1 B_2 \quad , \quad 
J^B_{-} =  \overline{B}^2 B_1 \quad , \quad 
J^B_3 = \overline{B}^1 B_1 - \overline{B}^2 B_2
\eea
where $J_{U(1)}$ is the current of the global axial $U(1)$ under which all chiral superfields have unit charge, whereas $J^A_i$ and $J^B_i$ are the currents associated to the generators of $SU(2)_A$ and $SU(2)_B$.
 
By direct inspection it is easy to obtain the $T^a_{i\overline{j}}$  coefficients in (\ref{eq:OPE}). Up to an overall $1/(4\pi)^{3}$, they are given by 
\bea
\label{T}
&& T^{U(1)}_{11} = (1-MN) \quad \quad T^{U(1)}_{22} = 2 (MN+1) \non
 \\
&& T^{U(1)}_{33} = T^{J^A_3}_{33} = T^{J^B_3}_{33} = 2 (MN+1) 
\quad \quad 
T^{U(1)}_{44} = -T^{J^A_3}_{44} = -T^{J^B_3}_{44} = 2 (MN+1)  
\non \\
&& T^{U(1)}_{55} = T^{J^A_3}_{55} = - T^{J^B_3}_{55} = 2 (MN+1) 
\quad \quad T^{U(1)}_{66} = -T^{J^A_3}_{66} =  T^{J^B_3}_{66} =  2 (MN+1)
\non \\
&& T^{U(1)}_{77} =   T^{J^A_3}_{77} = (MN+1)
\quad \quad T^{U(1)}_{88} = T^{J^B_3}_{88} = (MN+1) 
\non \\
&& T^{U(1)}_{99} = - T^{J^A_3}_{99} = (MN+1)
\quad \quad  T^{U(1)}_{10\,10} = - T^{J^B_3}_{10\,10} = (MN+1) \non\\
&&T^{J^A_{+}}_{27} = T^{J^A_{+}}_{4\, 10}=  T^{J^A_{+}}_{68}  =  T^{J^A_{+}}_{83} = T^{J^A_{+}}_{92} = T^{J^A_{+}}_{10\, 5} = 
2 (MN+1) \quad \quad 
\non\\&&
T^{J^B_{+}}_{28} = T^{J^B_{+}}_{49} = T^{J^B_{+}}_{57} = T^{J^B_{+}}_{73} = T^{J^B_{+}}_{96} = T^{J^B_{+}}_{10\, 2} = 2 (MN+1) \non\\&&
T^{J^A_{-}}_{ij} = T^{J^A_{+}}_{ji} \quad \quad T^{J^B_{-}}_{ij} = T^{J^B_{+}}_{ji}
\eea
Analogously, we can compute the OPE between two gauge currents. It has the general structure
\beq
J_M (x) J_N(0) \sim \frac{1}{|x|^3}   \G_{MN} k^a J_a + \cdots
\eeq
where $J_a$ are global currents. 

In the present case, taking the two gauge
currents $J_{A} = \bar{A}_a e^{V} A^a e^{-\hat{V}}$ and $J_B =  \bar{B}^a e^{V} B_a e^{-\hat{V}}$, the 
only nontrivial contribution at order $1/|x|^3$ is proportional to $J_{U(1)}$ with coefficient 
\beq
k^{U(1)} = -( 2MN + 1) \left( \frac{1}{K_1^2} + \frac{1}{K_2^2} \right) - 2 \frac{MN+2}{K_1 K_2} 
\label{kappa}
\eeq
This comes from two different contributions: The first one corresponds to expanding $J_A$ and $J_B$ 
at second order in the gauge prepotentials and contracting one chiral and 
two gauge superfields. The second contribution comes from normal--ordering the currents and corresponds to contracting 
two gauge superfields inside a single current with a one--loop corrected propagator.

Since the contraction of two gauge currents as well as of two chiral supermarginals give a logarithmically singular
contribution to the Kahler potential of the form $\int d^4 \theta Z^a(\mu) J_a$, 
the D--terms for the global symmetries have the general form \cite{Green:2010da}
\bea
D^a \equiv \mu \frac{\pa}{\pa \mu} Z^a(\mu)  \sim h^i\, T^a_{i\overline{j}}\, \overline{h^j} + k^a
\eea
Vanishing $D$-term conditions insure independence of the effective action of the energy scale and then
its superconformal invariance. 

In the present case,  using the explicit results (\ref{T}, \ref{kappa}) we  find
\bea
D_{U(1)} &\sim& (MN+1) \left(2 |h_2|^2 + 2 |h_3|^2 + 2 |h_4|^2 + 2 |h_5|^2 + 2 |h_6|^2 + |h_7|^2 + \right. \non\\&& \left. |h_8|^2 + |h_9|^2 + |h_{10}|^2 \right) - 2 (MN-1) |h_1|^2 + k^{U(1)} \non\\
D^A_{+} &\sim& 2 (MN+1) \left(h_2 \overline{h_7} + h_4 \overline{h_{10}} + h_9 \overline{h_2} + h_8 \overline{h_3} +  h_6 \overline{h_8} + h_{10} \overline{h_5} \right) \non\\
D^A_{-} &\sim& 2 (MN+1) \left( h_2 \overline{h_9} + h_3 \overline{h_8} + h_5 \overline{h_{10}} + h_7 \overline{h_2} + h_8 \overline{h_6} + h_{10} \overline{h_4} \right) \non\\
D^A_3 &\sim&  (MN+1)\left(2 |h_3|^2 - 2 |h_4|^2 + 2 |h_5|^2 - 2 |h_6|^2 + |h_7|^2 - |h_9|^2 \right) \non\\
D^B_{+} &\sim& 2 (MN+1) \left(h_2 \overline{h_8} + h_4 \overline{h_9} + h_5 \overline{h_7} + h_7 \overline{h_3} + h_9 \overline{h_6} + h_{10} \overline{h_2} \right) \non\\
D^B_{-} &\sim& 2 (MN+1) \left(h_2 \overline{h_{10}} + h_3 \overline{h_7} + h_6 \overline{h_9} + h_7 \overline{h_5} + h_8 \overline{h_2} + h_9 \overline{h_4} \right) \non\\
D^B_3 &\sim& (MN+1)\left(2 |h_3|^2 - 2 |h_4|^2 - 2 |h_5|^2 + 2 |h_6|^2 + |h_8|^2 - |h_{10}|^2 \right) 
\eea 
We recognize the $D_{\pm}=0$ conditions to be the same as  (\ref{eq:gamma}) for the vanishing of the off--diagonal part of the anomalous dimensions (\ref{eq:anomalous}), whereas the combinations $D_{U(1)} \pm D^A_3=0$ and $D_{U(1)} \pm D^B_3=0$ precisely match the conditions for the vanishing of the diagonal part of the anomalous dimensions, as derived above. This proves the complete equivalence of the two methods for finding the conformal manifold of the theory.

By using the nonconservation equation for the currents,
\bea
\bar{D}^2 J_a = h^i \, T^{a \ j}_{\ i} \, {\cal O}_j + \dots 
\eea
one can also check which operators are responsible for breaking the various symmetries of the free theory.\\
For example at the ABJ(M) fixed point, we see that the ABJM superpotential operator couples to the global axial $U(1)$ and thus becomes irrelevant, as we argued above.\\ At the ${\cal N} = 3$ fixed point, lying on the $SU(2)$ invariant subset of the conformal manifold, we see that indeed the conservation equations for $J^A_{+} - J^B_{-}$, $J^A_{-} - J^B_{+}$ and $J^A_3 - J^B_3$ are satisfied, corresponding to the preserved diagonal $SU(2)_D$ subgroup of $SU(2)\times SU(2)$. Instead, from the nonconservation of the currents of $(SU(2)\times SU(2))/SU(2)_D$, we see that the operators parameterized by $h_3-h_4$, $h_7 + h_{10}$ and $h_8 + h_{9}$ have become irrelevant by coupling to the broken currents.

 \section{Conformal manifold for flavored theories}\label{sec:4}

Finally, we study the conformal manifold for ${\cal N} = 2$ CS-matter theories when additional flavor degrees of freedom are included.

Flavor chiral superfields in the (anti)fundamental representation of the gauge groups can be safely added to the theories described above, without affecting superconformal invariance \cite{HK,GJ,HLT,Bianchi:2009rf}. For instance, given the model (\ref{action}, \ref{superpotential}) we can add $N_f$ fundamental and $\tilde{N_f}$ antifundamental  
chiral superfields $(Q_1^i,\tilde{Q}_{1,\,j})$, charged under the first gauge group and $M_f$ fundamentals and $\tilde{M_f}$ antifundamentals $(Q_2^k, \tilde{Q}_{2,\,l})$, charged under the second gauge group. In the literature only the cases $N_f =\tilde{N_f}$ and $M_f =\tilde{M_f}$ are discussed. However, as long as we are not concerned with surviving non--abelian global symmetries, nothing prevents us from considering a more general case with $N_f \neq \tilde{N_f}$, $M_f \neq \tilde{M_f}$.

For $W=0$, the classical global symmetry of the flavor sector is $U(N_f) \times U(\tilde{N_f}) \times U(M_f) \times U(\tilde{M_f})$ in addition to the $U(2) \times U(2)$ group of the bifundamental sector already discussed. 

The set of classical supermarginals is given by the operators belonging to the bifundamental sector (eqs. (\ref{superpotential}, \ref{marginal})) plus pure flavor operators
\beq
(F_{A,B})^{ik}_{\, \, \; jl}  \; \tilde{Q}_{A,\,i}\, Q_A^j\, \tilde{Q}_{B,\,k}\, Q_B^l \qquad , \qquad A,B = 1,2 ~{\rm (not ~summed})
\label{pure}
\eeq
with $(F_{A,B})^{ik}_{\, \, \; jl}  = (F_{B,A})^{ki}_{\, \, \; lj} $,
and mixed marginals whose form is constrained by gauge invariance to be 
\beq
M^{b , i}_{a , j}\;  \tilde{Q}_{1 \, i}  A^a B_b \, Q_1^j \qquad , \qquad       {\tilde M}^{b , i}_{a , j}  \;    \tilde{Q}_{2 \, i} B_b A^a \, Q_2^j
\label{mixed} 
\eeq
Counting the number of independent $M$ and ${\tilde M}$ couplings we have  $4(N_f \tilde{N_f} + M_f \tilde{M_f})$ operators
of the form (\ref{mixed}).  

For operators of the form (\ref{pure}), we have three different countings according to the values of the $A,B$ labels. For 
$A,B=1$, taking into account the symmetry of the trace, we have multiplicity $\frac12\, N_f \tilde{N_f} (N_f \tilde{N_f} + 1)$.
Analogously, for $A,B=2$ we have $\frac12\, M_f \tilde{M_f}  (M_f \tilde{M_f}  + 1)$  operators, whereas for $A=1$ and $B=2$
we find $N_f \tilde{N_f} M_f \tilde{M_f}$ supermarginals.

All together,  there are $ \frac12 (N_f \tilde{N_f} + M_f \tilde{M_f}) (N_f \tilde{N_f} + M_f \tilde{M_f} + 9) $ marginal operators in addition to the ten supermarginals of the unflavored case.
These operators break all the global symmetries except two diagonal $U(1)$'s among $U(N_f)\times U(\tilde{N_f})$ and $U(M_f)\times U(\tilde{M_f})$ which are part of the gauge symmetry. The number of broken global generators is then $ (N_f^2 + \tilde{N_f}^2 + M_f^2 + \tilde{M_f}^2 - 2) $. 

Using the arguments of \cite{Green:2010da,Kol:2010ub}, we find that in the presence of flavor degrees of freedom the dimension of the conformal manifold is  
\beq
{\rm dim}\, {\cal M}_c  = 5 + \frac12 (N_f \tilde{N_f} + M_f \tilde{M_f}) (N_f \tilde{N_f} + M_f \tilde{M_f} + 5) - (N_f - \tilde{N_f})^2 - (M_f - \tilde{M_f})^2  
\eeq
Setting $N_f = {\tilde N}_f$ and $M_f = {\tilde M}_f$ we obtain the dimension of the conformal manifold for the  
${\cal N}=3$ theory introduced in \cite{GJ} and studied in \cite{Bianchi:2009rf}.

\section{Conclusions}

For a large class of three dimensional, ${\cal N} = 2$ Chern--Simons matter theories, including the ${\cal N} = 6$ ABJ(M) and ${\cal N} = 3$ models, we have determined the structure of the conformal manifold, that is its dimension and the set of exactly marginal deformations. This has been accomplished first by applying general arguments based on global symmetry breaking \cite{Green:2010da,Kol:2010ub} and then by a direct 2--loop perturbative computation. In all the cases we have found perfect agreement between the two results, so proving that the method of \cite{Green:2010da,Kol:2010ub} can be consistently applied to three--dimensional models.\\
In the case of unflavored theories the conformal manifold turns out to be always a three--dimensional compact surface. Compactness seems to be a peculiar feature of three dimensional supersymmetric CFT's which does not have a direct analogous in four dimensions. The reason can be traced back to the fact that in this kind of theories the gauge couplings do not run.\\
We have extended our analysis to theories including fundamental flavor degrees of freedom. As expected, in this case the dimension of the conformal manifold turns out to be a function of the number of flavors.

\vskip 25pt
\section*{Acknowledgements}
\noindent 

This work has been supported in part by INFN and PRIN prot.20075ATT78-002.

\end{document}